\documentclass[a4paper]{article}

\usepackage{INTERSPEECH2021}
\usepackage{tablefootnote}
\usepackage{multirow,multicol}
\usepackage{caption}
\usepackage[symbol]{footmisc}
\usepackage{arydshln}

\newcommand\blfootnote[1]{%
  \begingroup
  \renewcommand\thefootnote{}\footnote{#1}%
  \addtocounter{footnote}{-1}%
  \endgroup
}

\title{Dual Script E2E Framework for Multilingual and Code-Switching ASR}
\name{Mari Ganesh Kumar, Jom Kuriakose, Anand Thyagachandran,  Arun Kumar A*, Ashish Seth*,  Lodagala V S V Durga Prasad*, Saish Jaiswal*, Anusha Prakash, and Hema A Murthy}
\address{
  Indian Institute of Technology Madras}
\email{mari@cse.iitm.ac.in (MGK), jom@cse.iitm.ac.in (JK), anushaprakash@smail.iitm.ac.in (AP), hema@cse.iitm.ac.in (HAM) }

\begin{document}

\maketitle

\begin{abstract}

  India is home to multiple languages, and training automatic speech recognition (ASR) systems is challenging. Over time, each language has adopted words from other languages, such as English, leading to code-mixing. Most Indian languages also have their own unique scripts, which poses a major limitation in training multilingual and code-switching ASR systems.
  
  Inspired by results in text-to-speech synthesis, in this paper, we use an in-house rule-based phoneme-level common label set (CLS) representation to train multilingual and code-switching ASR for Indian languages. We propose two end-to-end (E2E) ASR systems. In the first system, the E2E model is trained on the CLS representation, and we use a novel data-driven backend to recover the native language script. In the second system, we propose a modification to the E2E model, wherein the CLS representation and the native language characters are used simultaneously for training. We show our results on the multilingual and code-switching tasks of Indic ASR Challenge 2021. Our best results achieve $\approx 6\%$ and $5\%$ improvement in word error rate over the baseline system for the multilingual and code-switching tasks, respectively, on the challenge development data.
  
\end{abstract}

\section{Introduction}

    India is a multilingual country, having 22 official languages and many more mother tongues \cite{pattanayak1990multilingualism}. Building automatic speech recognition (ASR) systems for Indian languages is a challenging problem owing to the low resource setting. Moreover, due to globalization and migration, code-switching is a common phenomenon. Code-switching is the alternation between multiple languages in a single conversation. English is one of the most widely used languages in code-switching. The two major goals of the Indic ASR challenge 2021 \cite{Inter7329884:online, diwan2021multilingual} are the advancement of multilingual and code-switching ASR for Indian languages.
 \blfootnote{*authors contributed equally}   
     \par The main aim of this paper is to build multilingual ASR by pooling data from multiple low resource languages. We propose novel end-to-end (E2E) ASR systems using an in-house common label set (CLS) representation, which has been successful in text-to-speech (TTS) synthesis for low resource Indian languages \cite{Prakash2019_MLCM, Prakash20_Generic}. The proposed models are evaluated as part of the Indic ASR Challenge 2021.

    \par A major issue with building ASR systems for Indian languages in a multilingual context is that the languages have different grapheme representations. This difference in the grapheme representation limits the ASR models in learning common sounds across languages. Although Indian languages have different written scripts, they share a common phonetic base. Inspired by this, a common label set for 13 Indian languages was developed for TTS synthesis \cite{ramani}. In \cite{baby2016unified}, a rule-based unified parser for 13 Indian languages was proposed to convert grapheme-based Indian language text to phoneme-based CLS. CLS has shown to be successful for TTS systems developed using both hidden Markov model (HMM) based approaches \cite{ramani}, and recent end-to-end frameworks \cite{Prakash2019_MLCM, Prakash20_Generic}. For training multilingual ASR systems, recent studies use transliterated text pooled across various languages \cite{datta2020language, thomas2020transliteration}. In the context of multilingual ASR systems for Indian languages, \cite{vishwas-2021} used a simple character mapping based on CLS. The current work uses the phone-based CLS representation obtained using the rule-based unified parser \cite{baby2016unified} for training multilingual ASR systems.
    
    
    \par This paper uses the CLS representation in two different ways to train multilingual E2E transformer \cite{vaswani2017attention} models. First, we convert the native script of all Indian languages to the CLS representation and then train the E2E transformer model using the CLS representation. This model takes advantage of the same sounds across different languages and maps them together. This improves the performance of the system on low resource languages. However, the model decodes the audio files only in CLS. We propose novel language identification (LID) and machine transliteration techniques to recover the native script after decoding. This system achieves $5\%$ improvement over the baseline system for the multilingual task on the challenge development data.
    
    \par In the second system, we propose integrating the LID and machine transliteration backend within the E2E transformer model itself. We train this model using both the CLS representation and the native language script simultaneously. During decoding, the CLS output is discarded, and the model outputs the Indian language script directly. This system achieves a 6\% improvement over the baseline for the multilingual task on the challenge development data. We also show our results on the code-switching task of the challenge.  
    
    \par The remainder of the paper is organised as follows. The Indic ASR challenge and the associated datasets are briefly described in Section~\ref{sec:challenge_and_data}. All the ASR systems proposed in this paper use the CLS transcription for training. The conversion of native language script to the CLS representation is described in Section~\ref{sec:NStoCLS}. The proposed systems are explained in Section~\ref{sec:proposed_system}. Results are presented and discussed in Sections~\ref{sec:result} and ~\ref{sec:discussion}, respectively. The work is concluded in Section~\ref{sec:concl}.

\section{Indic ASR Challenge 2021}

\label{sec:challenge_and_data}

The Indic ASR challenge 2021 \cite{Inter7329884:online, diwan2021multilingual} consists of two sub-tasks. In sub-task 1, the main objective is to build a multilingual ASR system for Indian languages. The dataset for sub-task 1 consists of six low resource Indian languages, namely, Hindi, Marathi, Gujarati, Tamil, Telugu, and Odia. It is to be noted that, in the blind test, the spoken utterances are provided without any language information. The ASR system has to identify the language and decode the given utterance in the language-specific script. Of the six languages provided for this task, only Hindi and Marathi share the {\it ``Devanagari"} script, and the other four languages have their own script.

The goal of sub-task 2 is to build code-switching ASR systems. Hindi-English and Bengali-English are the two code-switched data pairs provided for this sub-task. Unlike sub-task 1, the language-pair information is provided in the blind set. However, the model should decode each word present in the code-switched utterance in the native script itself (Hindi/Bengali or English). The statistics of data provided for the two sub-tasks is presented in Table \ref{tab:Dataset}. For both the sub-tasks, the average WER computed across all languages/language pairs is used as the evaluation metric.

\begin{table}[h!t]
\centering
\caption{Datasets provided in the Indic ASR challenge}
\renewcommand{\arraystretch}{1.2}
\resizebox{\linewidth}{!}{
\begin{tabular}{c|c|c|c|c}
\hline
Language &
  Category &
  \begin{tabular}[c]{@{}c@{}} Duration\\ (Hrs) \end{tabular} &
  \begin{tabular}[c]{@{}c@{}}No. of\\ unique\\ sentences\end{tabular} &
  \begin{tabular}[c]{@{}c@{}}Average no.\\ of words\\ per sentence\end{tabular} \\ \hline
  	\multicolumn{5}{c}{Sub-Task 1} \\
    \hline
    \multirow{2}{*}{Telugu (te)}   & Train & 40 & 34,176 & 12.3 \\ \cline{2-5} 
                              & Dev & 5 & 2,997 & 9.4 \\ \hline
    \multirow{2}{*}{Tamil (ta)}    & Train & 40 & 30,239 & 7.3 \\ \cline{2-5} 
                              & Dev & 5 & 3,057 & 9.5 \\ \hline
    \multirow{2}{*}{Gujarati (gu)} & Train & 40 & 20,208 & 12.3 \\ \cline{2-5} 
                              & Dev  & 5 & 3,069 & 11.8 \\ \hline
    \multirow{2}{*}{Hindi (hi)}    & Train & 95.05 & 4,506 & 7.4 \\ \cline{2-5} 
                              & Dev & 5.55  &  386 & 12.2 \\ \hline
    \multirow{2}{*}{Marathi (mr)}  & Train & 93.89 & 2,543 & 6.8 \\ \cline{2-5} 
                              & Dev  & 5 & 200 & 6.7 \\ \hline
    \multirow{2}{*}{Odia (or)}     & Train & 94.54 & 820 & 9.1 \\ \cline{2-5} 
                              & Dev & 5.49  & 65 & 9.1 \\ \hline
    \multicolumn{5}{c}{Sub-Task 2} \\ \hline
    \multirow{2}{*}{Hindi-English (hi-en)}     & Train & 89.86 & 44244 & 12.0 \\ \cline{2-5} 
    & Dev & 5.18  & 2893 & 12.0 \\ \hline
    \multirow{2}{*}{Bengali-English (bn-en)}     & Train & 46.11 & 22386 & 9.4 \\ \cline{2-5} 
    & Dev & 7.01  & 3968 & 9.0\\ \hline
\end{tabular}
}
\label{tab:Dataset}
\end{table}

\section{Common Label Set (CLS) and Unified Parser}
\label{sec:NStoCLS}
    The first step in the proposed approach is to convert the text in terms of the CLS format \cite{ramani}. In CLS, similar phones across 13 Indian languages are mapped together and given a common label. Language-specific phones are given separate labels. The native text is converted to its constituent CLS representation using a unified parser \cite{baby2016unified}. The unified parser is a rule-based grapheme to phoneme converter designed for Indian languages. The unified parser takes a UTF-8 word as input and recognises the script based on the Unicode range. Then the parser applies the relevant language-specific rules and outputs the constituent phones in terms of CLS.
    
    \begin{figure}[h!t]
     \centering
     \captionof{table}{Examples of words and their corresponding CLS representations}
     \includegraphics[width = 0.8\linewidth]{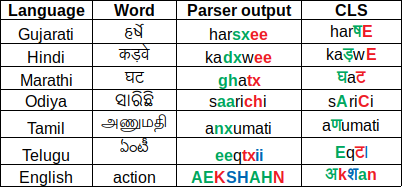}
     \label{fig:CLS}
    \end{figure}
    
    The CLS format uses a sequence of English characters for representation. Hence, to make the text purely phonetic, each CLS label is further mapped to a single character. For example, the CLS label ``aa'' (long vowel ``a'', as in {\it art}) is mapped to character ``A'', CLS label ``ph'' (aspirated stop consonant ``p'') to character ``P'', and so on. This CLS-like mapping was proposed in \cite{Prakash2019_MLCM}. The word to final CLS mapping is illustrated by examples from different languages in Table \ref{fig:CLS}.

  In the code-switching task, English words are also present in the text. English words are first parsed through a neural network-based grapheme to phoneme converter \cite{g2pE2019}, and the constituent phones are obtained in terms of the Carnegie Mellon University (CMU) phone set. These labels are then mapped to CLS representation as proposed in \cite{AnjuIS18}. An example of the word ``action'', its parsed output in CMU phones and the final CLS representation is given in Table \ref{fig:CLS}.

\section{Multilingual ASR Systems}
\label{sec:proposed_system}
The audio files were first downsampled to 8000Hz, and  80 mel filter bank energies along with pitch were extracted to be used as features. 
The paper uses three different E2E models (1 baseline, 2 proposed) for building multilingual ASR.  The three variants of the E2E model are discussed in this section. ESPNet toolkit \cite{watanabe2018espnet} was used to train the E2E models. The models were trained using Hybrid CTC-attention \cite{watanabe2018espnet, watanabe2017hybrid} based transformer architectures.  

\begin{figure*}[h!t]
\centering
\includegraphics[width =\linewidth]{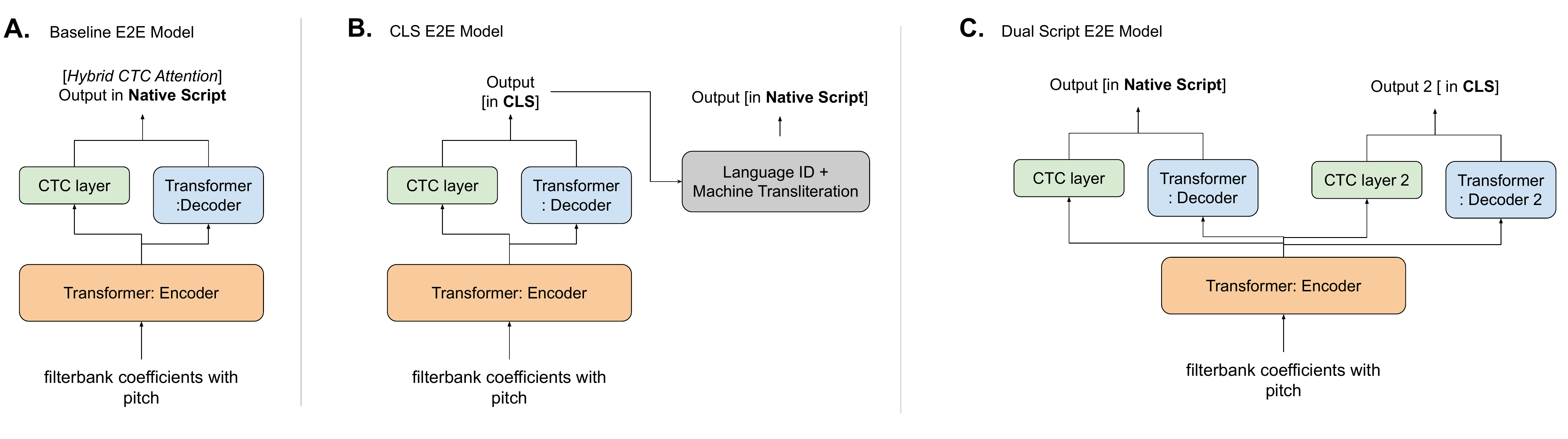}
\caption{Illustration of proposed E2E hybrid CTC-attention models using CLS}
\label{fig:E2EModels}
\end{figure*}

\subsection{Baseline E2E Model}
\label{sec:base2e}
This simple baseline model pools data from all the languages during training to build a multilingual system. Since different Indian languages use different scripts, this model does not take advantage of the common sounds present across all the languages.
\subsection{CLS E2E Model}
\label{sec:clse2e}
\begin{figure}[h!t]
     \centering
     \captionof{table}{Confusions in CLS to native script mapping}
     \includegraphics[width = 0.7\linewidth]{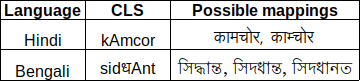}
     \label{tab:cls_2_unicode}
\end{figure}

Instead of the native language script, the CLS E2E model uses a common representation obtained using the unified parser (see Section~\ref{sec:NStoCLS}).  Since this model is trained using CLS, it decodes back only in CLS labels. It is to be noted that the native text (in UTF-8) to CLS mapping is not one-to-one due to rules such as \textit{schwa} deletion, geminate correction, and syllable parsing \cite{baby2016unified}. Due to these different rules, one-to-one mapping of CLS back to the native script of the right language is challenging. Examples of such confusions are given in Table~\ref{tab:cls_2_unicode}.


Since the native script cannot be retrieved from the CLS using a one-to-one mapping, machine transliteration models are trained and used.  Machine translation techniques used for language translation are extended to transliteration by inserting space between each character in a word and considering each character as a word. Neural machine transliteration models are trained using the default configuration of open neural machine translation (ONMT) toolkit \cite{klein2017opennmt} to learn the character level mapping from CLS to the native text. A multinomial naive Bayes classifier using TF-IDF features obtained from decoded CLS is used to identify the language.

\subsection{Dual Script E2E Model}
\label{sec:dual2e}
In this system, we integrate the LID and machine transliteration backend within the E2E model by adding additional decoders. We use two CTC layers and decoders to predict the CLS  and native language script (in UTF-8) simultaneously. This modified hybrid CTC-attention model is trained over a loss function, giving equal weight to predicting both the CLS and native language script. This simultaneous training is expected to cue the E2E model with phonemes that are common across languages.  During decoding, the CTC layer and the decoder corresponding to CLS are discarded, and the model outputs the Indian language script directly. Figure~\ref{fig:E2EModels} illustrates the differences between the three E2E models trained in this work.

\subsection{Implementation Details}

\par For all the three different E2E models mentioned above, the same network architecture was used. Table \ref{tab:network_config} presents the network architecture in detail. These E2E models were trained with both byte-pair sub-word units (BPU) \cite{kudo2018subword} and character units (CU). For sub-task 1, the E2E models were trained only using the corresponding sub-task data for 20 epochs. For sub-task 2, the models obtained in sub-task 1 were fine-tuned for 20 epochs. No external data was used for training acoustic models.

\begin{table}
\centering
\caption{Detailed transformer architecture used for E2E acoustic models}
\label{tab:network_config}
\resizebox{0.9\linewidth}{!}{
\begin{tabular}{ll|ll|ll} 
\toprule
Name                                                          & Value & Name                                                                                                & Value~ & Name                                                       & Value  \\ 
\hline
\begin{tabular}[c]{@{}l@{}}Attention\\Dimensions\end{tabular} & 256   & \begin{tabular}[c]{@{}l@{}}Encoder\\Layers\end{tabular}                                             & 12     & \begin{tabular}[c]{@{}l@{}}Decoder\\Layers\end{tabular}    & 6      \\ 
\hdashline[1pt/1pt]
\begin{tabular}[c]{@{}l@{}}Attention\\Heads\end{tabular}      & 4     & \begin{tabular}[c]{@{}l@{}}Encoder\\Dimension\end{tabular}                                          & 2048   & \begin{tabular}[c]{@{}l@{}}Decoder\\Dimension\end{tabular} & 2048   \\ 
\hdashline[1pt/1pt]
\begin{tabular}[c]{@{}l@{}}Output\\BPU\end{tabular}           & 5000  & \begin{tabular}[c]{@{}l@{}}Output 2\\BPU \tablefootnote{Used for output 2 (see Figure 1. C) in dual script model}\end{tabular} & 800    &                                                            &        \\
\bottomrule
\end{tabular}
}
\end{table}

Additionally, for sub-task 1, a transformer-based language model was trained using external text data. Text data from Indic TTS \cite{ArunResources2016}, and AI4Bharat NLP corpora \cite{kakwani2020indicnlpsuite}  were used to train the language model. The transformer-based \cite{vaswani2017attention} language model was trained for 1 epoch on the combined text corpora with about 150 million sentences pooled from all six languages used in sub-task 1. No language model was used for sub-task 2.

For the CLS E2E model, long short term memory (LSTM)  based encoder-decoder model with global attention was used for neural machine transliteration at the word level (see Section~\ref{sec:clse2e}). The machine transliteration models were trained language-wise. Since the character vocabulary of CLS to the native script is limited, the system learns transliteration. When trained with the IndicTTS text data \cite{ArunResources2016} and the ASR Challange train text, this model achieved 1.78\% WER and 0.44\% character error rate (CER) on sub-task 1 development data, averaged across all six languages.

For the LID part of the CLS E2E model, TF-IDF features were used. Term-frequency (TF) gives more importance to a more frequent word in a document, whereas inverse-document frequency (IDF) tries to penalize a word if it occurs in multiple documents. Multi-gram TF-IDF features were extracted at character and word levels from each sentence, and a simple Bayes classifier was used to predict the language. This system was also trained with IndicTTS text data \cite{ArunResources2016} and ASR Challange train text, and achieved an accuracy of $99.7\%$ on sub-task 1 development data.

For sub-task 1,  HMM-based and time delay neural network (TDNN) based models were provided as baseline models. For sub-task 2, in addition to these models, an E2E conformer \cite{gulati2020conformer} model was also used as a baseline. The details of these baseline models can be found in \cite{diwan2021multilingual}.

\section{Results}
\label{sec:result}

\subsection{Results of sub-task 1}
\label{sec:st1}
Table~\ref{tab:subtask1:dev} shows the results of all the models discussed in this paper on the development data for sub-task 1. The dual script system, which uses the character units (System H), outperforms the best baseline (System B) by 1\% average WER, without using any language model. With a language model, this system (System L) outperforms the best baseline by 6\% average WER.

\begin{table}
\centering
\caption{Results of sub-task 1 on  development data}
\label{tab:subtask1:dev}
\renewcommand{\arraystretch}{1.2}
\resizebox{\linewidth}{!}{
\begin{tabular}{c|l|c|c|c|c|c|c|c|c} 
\toprule
\begin{tabular}[c]{@{}c@{}}System\\ID\end{tabular} & \multicolumn{1}{c|}{\begin{tabular}[c]{@{}c@{}}System\\Type\end{tabular}}        & \begin{tabular}[c]{@{}c@{}}BPU/\\CU\end{tabular} & hi            & mr            & or            & ta            & te            & gu            & Avg            \\ 
\hline
\multicolumn{10}{c}{Baseline Methods (provided with Indic ASR Challenge)}                                                                                                                                                                                                                                 \\ 
\hline
A                                                  & GMM-HMM                                                                          & -                                                & 69.0          & 33.2          & 55.7          & 48.8          & 47.2          & 28.3          & 46.8           \\ 
\hline
B                                                  & TDNN                                                                             & -                                                & 40.4          & 22.4          & 39.0          & 33.5          & 30.6          & 19.2          & 30.7           \\ 
\hline
\multicolumn{10}{c}{Results without any Language Model}                                                                                                                                                                                                                                                   \\ 
\hline
C                                                  & \multirow{2}{*}{\begin{tabular}[c]{@{}l@{}}Baseline \\E2E Model\end{tabular}}    & BPU                                              & 52.1          & 33.8          & 71.3          & 31.3          & 32.9          & 26.5          & 49.5           \\ 
\cline{1-1}\cline{3-10}
D                                                  &                                                                                  & CU                                               & 26.5          & \textbf{17.1} & \textbf{36.1} & 35.3          & 36.6          & 28.4          & 30.0           \\ 
\hline
E                                                  & \multirow{2}{*}{\begin{tabular}[c]{@{}l@{}}CLS \\E2E Model\end{tabular}}         & BPU                                              & 34            & 21.8          & 50.1          & 31.7          & 31.5          & 26.5          & 32.6           \\ 
\cline{1-1}\cline{3-10}
F                                                  &                                                                                  & CU                                               & 26.2          & 17.4          & 39.5          & 37.8          & 37.2          & 30.1          & 34.6           \\ 
\hline
G                                                  & \multirow{2}{*}{\begin{tabular}[c]{@{}l@{}}Dual Script \\E2E Model\end{tabular}} & BPU                                              & 29.4          & 19.8          & 44.9          & \textbf{30.5} & \textbf{31.9} & \textbf{24.4} & 30.1           \\ 
\cline{1-1}\cline{3-10}
H                                                  &                                                                                  & CU                                               & \textbf{25.9} & \textbf{17.1} & 37.4          & 35.2          & 35.8          & 27.7          & \textbf{29.8}  \\ 
\hline
\multicolumn{10}{c}{Results with Language Model}                                                                                                                                                                                                                                                          \\ 
\hline
I                                                  & \multirow{2}{*}{\begin{tabular}[c]{@{}l@{}}CLS \\E2E Model\end{tabular}}         & BPU                                              & 31.8          & 21.8          & 48.2          & 25.6          & 24.2          & 20.7          & 28.7           \\ 
\cline{1-1}\cline{3-10}
J                                                  &                                                                                  & CU                                               & \textbf{21.4} & \textbf{14.6} & 38.3          & 28.8          & 27.3          & 22.4          & 25.4           \\ 
\hline
K                                                  & \multirow{2}{*}{\begin{tabular}[c]{@{}l@{}}Dual Script \\E2E Model\end{tabular}} & BPU                                              & 27.8          & 20.0          & 48.2          & \textbf{23.6} & \textbf{23.6} & \textbf{18.8} & 27.0           \\ 
\cline{1-1}\cline{3-10}
L                                                  &                                                                                  & CU                                               & 21.6          & \textbf{15.1} & \textbf{36.0} & 25.9          & 25.3          & 20.5          & \textbf{24.0}  \\
\bottomrule
\end{tabular}
}
\end{table}

\begin{table}
\centering
\caption{Results of sub-task 1 on blind data}
\label{tab:subtask1:blind}
\renewcommand{\arraystretch}{1.2}
\resizebox{0.88\linewidth}{!}{
\begin{tabular}{c|c|c|c|c|c|c|c|c} 
\toprule
\begin{tabular}[c]{@{}c@{}}System\\ID\end{tabular} & hi            & mr            & or            & ta            & te            & gu            & Avg-1          & Avg-2          \\ 
\hline
\multicolumn{9}{c}{Baseline}                                                                                                                                                         \\ 
\hline
B                                                  & 37.2          & \textbf{29.0} & 38.4          & 34.0          & 31.4          & 26.1          & \textbf{32.73} & 33.4           \\ 
\hline
\multicolumn{9}{c}{Final Systems Submitted}                                                                                                                                          \\ 
\hline
J                                                  & 19.5          & 85.9          & 37.1          & 32.0          & 30.3          & 32.9          & 39.6           & 30.3           \\ 
\hline
K                                                  & 25.3          & 100.3         & 51.2          & \textbf{25.1} & \textbf{25.4} & \textbf{25.4} & 42.1           & 30.4           \\ 
\hline
L                                                  & \textbf{17.8} & 111.7         & \textbf{32.1} & 27.1          & 28.1          & 29.8          & 41.1           & \textbf{27.1}  \\
\bottomrule
\end{tabular}
}
\end{table}

The results of the final systems submitted to the challenge on the blind test set are given in Table~\ref{tab:subtask1:blind}. In contrast to results in Table~\ref{tab:subtask1:dev}, for the blind data, the baseline system  gave better results than the proposed systems (in Avg-1, computed from all 6 languages). However, it is notable that, without Marathi, the average WER (Avg-2) improved by 6\% over the best baseline for the dual script system (System L). These results are discussed in detail in Section~\ref{subsec:CLS}.

\subsection{Results of sub-task 2}
\label{sec:st2}
Table~\ref{tab:subtask2} shows the sub-task 2 development and blind data results for all the models discussed in this paper. Unlike results in Section~\ref{sec:st1}, the byte-pair models are observed to give better results than the character-level models. The dual script system with byte-pairs (System Q) gave $\approx$ 5\% improvement in WER over the baseline system (System O) on both development and blind data sets. Also, we highlight that this system was trained with combined Hindi-English and Bengali-English data, and the E2E model predicts the language pair during decoding (similar to sub-task 1). In contrast, the baseline system was trained separately for each language pair and decoded with language-pair information.

\begin{table}[h!t]
\centering
\caption{Results of sub-task 2 on development and blind data}
\label{tab:subtask2}
\renewcommand{\arraystretch}{1.2}
\resizebox{\linewidth}{!}{
\begin{tabular}{c|c|c|c|c|c|c|c|c} 
\toprule
\multirow{2}{*}{\begin{tabular}[c]{@{}c@{}}System\\ID\end{tabular}} & \multirow{2}{*}{\begin{tabular}[c]{@{}c@{}}System\\Type\end{tabular}}            & \multirow{2}{*}{\begin{tabular}[c]{@{}c@{}}BPU/\\CU\end{tabular}} & \multicolumn{3}{c|}{Dev Data} & \multicolumn{3}{c}{Blind Test}  \\ 
\cline{4-9}
                                                                    &                                                                                  &                                                                   & hi-en & bn-en         & Avg   & hi-en         & bn-en & Avg     \\ 
\hline
\multicolumn{9}{c}{Baseline Methods (provided with Indic ASR Challenge)}                                                                                                                                                                                                                     \\ 
\hline
M                                                                   & GMM-HMM                                                                          & -                                                                 & 44.3  & 39.1          & 41.7  & -             & -     & -       \\ 
\hline
N                                                                   & TDNN                                                                             & -                                                                 & 36.9  & 34.0          & 35.6  & -             & -     & -       \\ 
\hline
O                                                                   & E2E Model                                                                        & BPU                                                               & 27.7  & 37.2          & 32.4  & 25.5          & 32.8  & 29.1    \\ 
\hline
\multicolumn{9}{c}{Results without any Language Model}                                                                                                                                                                                                                                       \\ 
\hline
P                                                                   & \multirow{2}{*}{\begin{tabular}[c]{@{}c@{}}Dual Script \\E2E Model\end{tabular}} & CU                                                                & 33.0  & 27.0          & 30    & -             & -     & -       \\ 
\cline{1-1}\cline{3-9}
Q                                                                   &                                                                                  & BPU                                                               & 28.9  & \textbf{25.3} & 27.1  & \textbf{22.0} & 27.8  & 24.9    \\
\bottomrule
\end{tabular}
}
\end{table}

\section{Discussion}
\label{sec:discussion}

\subsection{CLS for E2E ASR system}
\label{subsec:CLS}
In this paper, we propose two novel multilingual ASR systems that are trained with CLS. From the results in Tables~\ref{tab:subtask1:dev}~and~\ref{tab:subtask2}, it is observed that E2E models give better results when trained with CLS instead of just pooling the native language script. This observation is common to both E2E systems trained with byte-pair and character units. This observation verifies our conjecture that CLS cues the E2E model in learning common sounds across languages, thus enabling faster convergence in a low resource setting.

In Tables~\ref{tab:subtask1:dev}~and~\ref{tab:subtask2}, we see that the dual script system consistently outperforms the CLS system with a separate backend. However, in the blind set of sub-task 1 (Table~\ref{tab:subtask1:blind}), we observe that the CLS system with a separate backend generalises better, especially for Marathi. The CLS model with a separate backend is able to scale better mainly because of the very few unique sentences in the training data for Marathi (see Table~\ref{tab:Dataset}) and usage of more text data to train the backend (see Section~\ref{sec:proposed_system}). We also note that the Indic ASR challenge website \cite{Inter7329884:online} mentions that the channel encoding scheme between the Marathi blind and non-blind (train \& test) datasets are different.

\subsection{Byte-pair and character units for low resource languages}

From the results in Table~\ref{tab:subtask1:dev}, it is observed that the byte-pair system consistently gives better results for Tamil, Telugu and Gujarati. In comparison, the character level system achieves better results for Hindi, Marathi and Odia. This is mainly because the training data of Hindi, Marathi and Odia has many repeated sentences. This has led to poor byte-pair estimation compared to the other three languages, which have about 30,000 unique sentences.

\subsection{Limitations}

Although sub-task 1 is monolingual data, we observe that Marathi and Odia have many English words in the train and development sets, with very few unique and short sentences (see Table~\ref{tab:Dataset}). Also, the code-switched data in sub-task 2 is very noisy. It contains English words in the Indian language script with many additional characters such as Greek letters \cite{Inter7329884:online}.

\section{Conclusion}
\label{sec:concl}

Given the low resource setting and multiple grapheme representations of Indian languages, training ASR systems is challenging. The use of CLS for training multilingual E2E ASR systems ensures that common sounds are mapped together, leading to better ASR models.  We also propose a dual script architecture trained simultaneously on CLS and native language text. This model can be used with a rule-based many-to-one mapping for different scripts. The model learns the native script and also gets the advantage of the common labels. Results across various models also show that the choice of basic units (CU or BPU) largely depends on the available training data.

\section{Acknowledgements}

This work was carried out as a part of the following projects-- ``Natural Language Translation Mission'' (CS2021012MEIT003119) and ``Speech to Speech Machine Translation'' (CS2021152OPSA003119), funded by the Ministry of Electronics and Information Technology (MeitY) and the Office of the Principal Scientific Advisor (PSA) to the Government of India, respectively.

\bibliographystyle{IEEEtran}

\bibliography{mybib}


\end{document}